\documentclass[twocolumn,aps,prl,amsmath,amssymb,floatfix,superscriptaddress]{revtex4}
\usepackage{times,xspace}

\usepackage{graphicx}% Include figure files
\usepackage{dcolumn}% Align table columns on decimal point
\usepackage{bm}% bold math
\usepackage{amssymb}

\begin{document}

%\preprint{Preprint}

%\title{Monopole polarons and molecules in the pentagonal spin ice}
\title{Magnetic monopole polarons in spin ice with mixed coordination numbers}

\author{Gia-Wei Chern}
\affiliation{Center for Nonlinear Studies and Theoretical Division, Los Alamos National Laboratory, Los Alamos, NM 87545, USA}

\author{Paula Mellado}
\affiliation{Department of Engineering and Sciences, Universidad Adolfo Iba\~nez, Chile}

\date{\today}

\begin{abstract}
Emergent quasiparticles that arise from the fractionalization of the microscopic degrees of freedom have
been one of the central themes in modern condensed matter physics. The notion of magnetic monopoles, freely moving quasiparticles fragmented from
local dipole excitations, has enjoyed much success in understanding the thermodynamic, static, 
and transport properties of the so-called spin-ice materials.
The artificial version of spin ice, where a lattice of nanoscale magnetic dipoles is sculpted out of a ferromagnetic film, provides
a unique opportunity to study these unusual quasiparticles in a material-by-design approach. 
Here we show that the elementary excitations in the ice phase of a nano-magnetic array arranged in the pentagonal lattice
are  composite objects comprised of the emergent monopole and a surrounding cloud of opposite 
uncompensated magnetic charges. 
\end{abstract}

\maketitle

Spin-ice materials~\cite{bramwell01,castelnovo12} such as Ho$_2$Ti$_2$O$_7$ and Dy$_2$Ti$_2$O$_7$ are a class of geometrically 
frustrated ferromagnets that retain an extensive residual entropy even at very low temperatures~\cite{ramirez99}. 
The magnetic ions in these compounds sit on a network of corner-sharing
tetrahedra that is commonly referred to as the pyrochlore lattice. The strong on-site crystal field forces the atomic moment to point in or out of the tetrahedra.
In conjunction with the dipolar interactions, this uniaxial anisotropy gives rise to the two-in-two-out ice rules, similar to the Bernal-Fowler rules in water ice, 
that dictate local ordering of spins in a tetrahedron. 
The nonzero residual entropy density in spin ice originates from the fact that an extensively large number of spin configurations satisfy
the 2-in-2-out ice rules in a macroscopic sample. Inverting a local moment in the ice phase creates two defect tetrahedra that violate the ice rules. 
It was shown that these defect tetrahedra are deconfined quasiparticles carrying a nonzero magnetic charge~\cite{castelnovo08,castelnovo11,jaubert09}.
The low-energy excitations in spin ice can thus be viewed as a weakly interacting plasma of emergent magnetic monopoles.

A two-dimensional analogue of the pyrochlore spin ice has been realized in arrays of interacting single-domain ferromagnetic
nanoislands arranged as links of a square lattice~\cite{wang06,nisoli07,morgan11a}. The basic units in this artificial square ice are vertices at which four islands meet;
they correspond to the tetrahedra in the pyrochlore network. Similar to the pyrochlore spin ice, 
the lowest-energy vertices in square ice are also magnetically charge neutral with two of the
moments pointing in toward the vertex, and two pointing out. 
A rather different picture emerges in spin ices composed of vertices with odd coordination numbers. A well studied example is the kagome ice 
which can be realized in a hexagonal network of nanomagnetic islands~\cite{tanaka06,qi08,ladak10,mengotti11}. 
The ice rules in this case require that each vertex has two spins
pointing inward and one outward or vice versa, leaving a net magnetic charge at each vertex.
The existence of these uncompensated magnetic charges leads to a novel classical spin liquid phase in kagome ice:
the residual magnetic charges develop a NaCl-type staggered ordering while the magnetic moments remain disordered~\cite{moller09,chern11,rougemaille11,branford12}.

\begin{figure*}
\includegraphics[width=1.75\columnwidth]{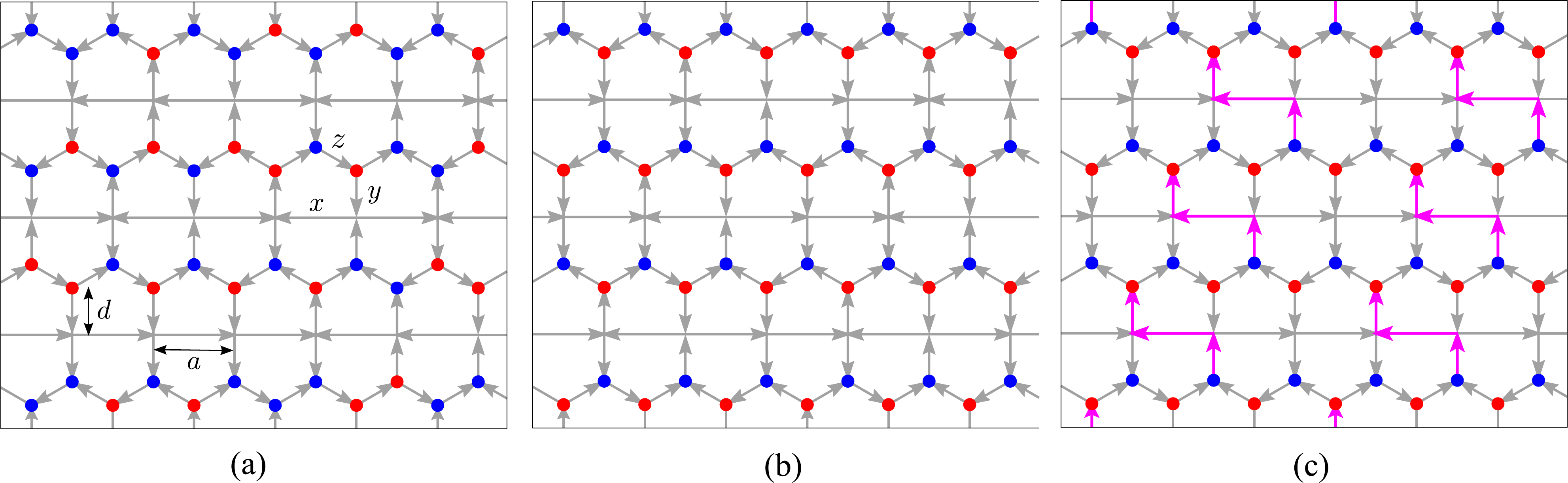}
\caption{\label{fig:lattice}  Pentagonal spin ice. (a) A spin-ice microstate without charge nor spin order; the ice rules are satisfied at all vertices.
The blue and red circles denote $Q = +1$ and $-1$ vertices, respectively.
(b) A charge-ordered spin-ice microstate lacking the spin order. (c) One of 
the dipolar ordered spin-ice states.
}
\end{figure*}

Motivated by the intriguing interplay between emergent monopoles and the uncompensated charges  residing at odd vertices, 
in this paper we consider dipolar magnetic arrays arranged in the so called pentagonal network~\cite{waldor,moessner01} which contains
both even and odd vertices; see Fig.~\ref{fig:lattice}. Remarkably,
we find that the magnetic Coulomb interactions in the pentagonal spin ice gives rise to a new type of quasiparticles: the magnetic monopole polarons.
%The even vertices in this network constrained by the ice rules are magnetically
%charge neutral at low temperatures, whereas nonzero magnetic charges remain at vertices with odd coordination number. 
Due to the presence of both even and odd vertices,
thermally excited monopoles that violate the 2-in-2-out ice rules at even vertices interact with a background of uncompensated charges at the odd vertices. 
Through the magnetic Coulomb interactions, a nearby cloud of opposite charges is created by the emergent monopoles, forming a composite
quasiparticle similar to the electron polarons in a crystalline solid~\cite{landau33}.
While the pentagonal spin ice is probably the simplest system exhibiting this novel monopole-polaron ice phase, 
we believe that the existence of such composite quasiparticles is a general feature in spin ices with mixed coordination numbers.

The pentagonal lattice shown in Fig.~\ref{fig:lattice} can be viewed as an elongated honeycomb lattice
with additional links added to produce even vertices (with coordination number $z=4$). A background of residual magnetic charges
reside on the $z=3$ vertices of the original honeycomb lattice. For convenience, we classify the links
into three categories: $x$ and $y$ refer to the two types of links that are perpendicular to each other,
while type-$z$ denotes the links on the zigzag chain of the lattice; see Fig.~\ref{fig:lattice}(a).
We choose a lattice parameter $d = a/\sqrt{3}$, where $a$ is the lattice constant, 
such that the $C_3$ rotational symmetry is preserved at the $z=3$ sites.

The interactions between the dipolar needles or nano-islands are described by the Hamiltonian
\begin{equation}
	\label{eq:H_dip}
	H = \frac{\mu_0}{4\pi}\! \int \! d\mathbf r_1 d\mathbf r_2 \frac{\mathbf  m(\mathbf r_1)\!\cdot\!\mathbf m(\mathbf r_2) 
	- 3 [\mathbf m(\mathbf r_1)\cdot\hat{\mathbf r}_{12}] [\mathbf m(\mathbf r_2)\cdot\hat{\mathbf r}_{12}]}{ |\mathbf r_{12}|^3},
\end{equation}
where $\mathbf r_{12} = \mathbf r_1 - \mathbf r_2$, and $\mathbf m(\mathbf r)$ is the dipole moment at $\mathbf r$,
which points along the direction of the link. 
Minimization of the dominant nearest-neighbor dipolar interactions gives rise to the ice rules~\cite{bramwell01,castelnovo12}.
The energetics associated with the ice rules and the long-range interactions contained in Eq.~(\ref{eq:H_dip}) 
can be intuitively understood by introducing an effective magnetic charge $Q_\alpha$ defined as the number of spins 
pointing inward minus the number of spins pointing outward at each vertex $\alpha$.
The ice rules dictate that locally the lowest-energy vertices are those with the minimum magnetic charge.
Consequently, the lowest-energy $z=3$ vertices have a total charge $Q = \pm 1$ (1-in-2-out or 2-in-1-out),
and the excited states are vertices with $Q=\pm 3$ (all-in or all-out).

For the $z=4$ vertices, the lowest-energy state corresponds to the 2-in-2-out configurations with a zero magnetic charge.
Unlike the three-dimensional pyrochlore spin ice, the planar geometry in general results in an inequivalence between the six 2-in-2-out configurations:
two of them, labeled Type-I in Ref.~\onlinecite{wang06},  have zero total magnetic moment, while the others (Type-II) have a net moment along a diagonal direction.
Moreover, the difference in the length of $x$ and $y$ links in the pentagonal lattice also gives rise to two inequivalent excited vertices with a charge $Q=\pm 2$. 
These high-energy vertices have either 3-in-1-out or 1-in-3-out spin configurations; their energy depends on whether
the minority spin, e.g. the spin pointing outward in a 3-in-1-out vertex, is on the $x$ or $y$ links. 
At the level of nearest-neighbor dipolar interactions, both inequivalences can be remedied by introducing a height displacement $h$ between
the $x$ and $y$ nano-islands~\cite{moller06,mol10} and by adjusting the magnetic moment on the $x$-links. 
By treating the dipoles as point-like objects $\mathbf m(\mathbf r) = \bm\mu_i \delta(\mathbf r - \mathbf r_i)$, the on-site energies $\varepsilon_{|Q|} $
of vertices with a total charge $Q$ are: $\varepsilon_{0} = -4D$, $\varepsilon_{1} = -(14/3\sqrt{3})D$, $\varepsilon_{2} = 0$, and $\varepsilon_{3} = +(14/\sqrt{3})D$.
Here $D = \frac{\mu_0}{4\pi}\frac{\mu^2}{d^3}$, $\mu$ is the dipole moment on the $y$ and $z$ links, while the moment on the $x$-links is $3^{3/4}\mu$,
and the height offset is $h/d = \bigl[\bigl(3\sqrt{3}/4\bigr)^{3/5} - 1\bigr]^{1/2}$.

\begin{figure}
\includegraphics[width=0.84\columnwidth]{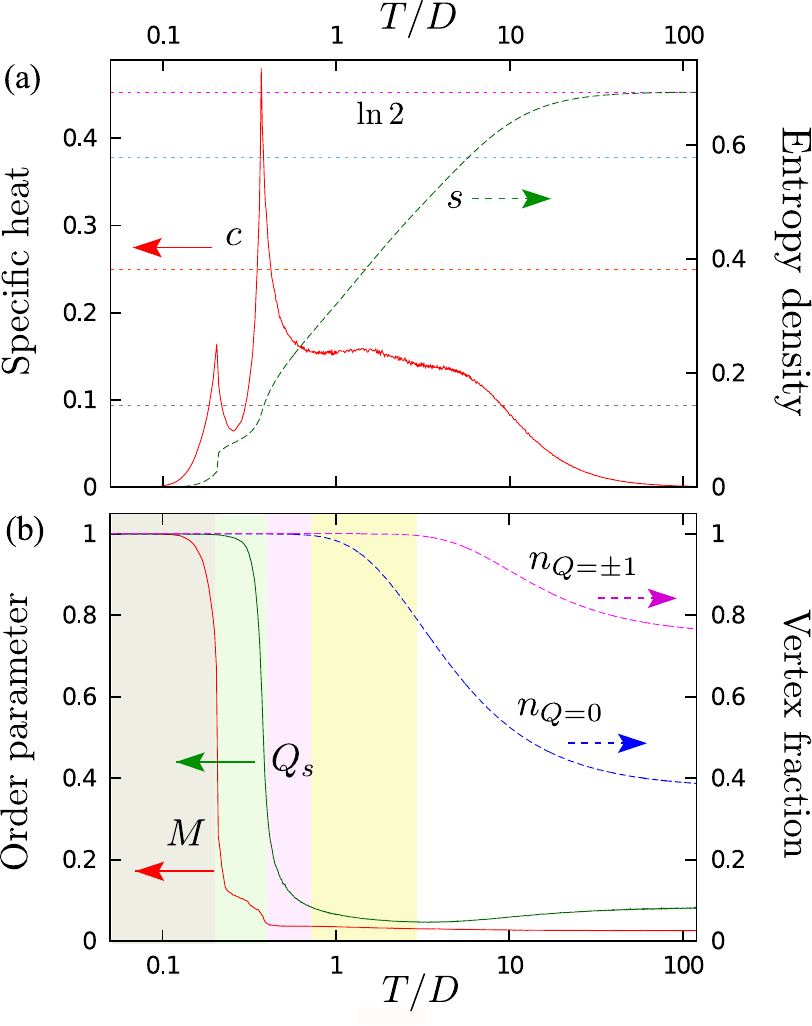}
\caption{\label{fig:simulation} (a) Temperature dependence of the specific heat $c(T)$ (left axis) and entropy per spin $s(T)$ (right axis)
of the pentagonal spin ice using the point-dipole approximation. The horizontal dashed lines denote the entropy density
of the high-$T$ Ising paramagnetic phase $\ln 2\approx 0.693$, and the three ice phases: $s_{z3} =  \frac{2}{5}\ln 3\sqrt{2} \approx 0.578$,
$s_{z4} =  \frac{1}{5}\ln \frac{27}{4} \approx 0.3819$, and $s_{\rm co} \approx 0.143$. (b) Temperature dependence of the
magnetic $M$ and charge $Q_s$ order parameters (left axis) and the fraction of vertices $n$ satisfying the ice rules (right axis).}
\end{figure}

The presence of two sets of ice rules associated with the odd and even vertices, respectively, indicates two distinct ice phases in the pentagonal lattice.
%Since the self-energy of on-site magnetic charge is proportional to $Q^2$,  the energy scale of suppressing the $Q = \pm 3$ charges at the $z=3$ vertices
%is generally greater than that of ice rules for the even coordination number sites.  
Thermodynamically, the system undergoes two successive crossovers into these ice phases upon lowering the temperature.
The crossover from the high-temperature paramagnetic regime to the ice-I phase  [yellow sector in Fig.~\ref{fig:simulation}(b)]
is marked by a broad peak in the specific heat, as shown in Fig.~\ref{fig:simulation}(a).
The numerical $c(T)$ curve is obtained from Monte Carlo simulations with full dipolar interactions and periodic boundary conditions 
on a lattice of $N = 5\times 15^2$ spins. As Fig.~\ref{fig:simulation}(b) shows, the fraction of vertices satisfying ice rules $Q = \pm 1$ 
saturates to 1 below $T_{z3} \sim \Delta_{3} \approx 15.1 D$, where $\Delta_{3} = (\varepsilon_{3} - \varepsilon_{1}) + \delta_3 $,
and $\delta_3$ is the correction due to the long-range part of the dipolar interactions~\cite{castelnovo08}.

The crossover into the ice-II phase [cyan sector in Fig.~\ref{fig:simulation}(b)], indicated by a second broad peak 
in the $c(T)$ curve [Fig.~\ref{fig:simulation}(a)], corresponds to the suppression of the $Q=2$ topological defects (or monopoles) at the even vertices.
Interestingly, contrary to the $Q= \pm 3$ excitations at the $z=3$ sites, we find a monopole excitation energy $\Delta_2 \approx \varepsilon_2 - \varepsilon_0$
which does {\em not} acquire a significant correction $\delta_2$ from the long-range dipolar interactions.
As we discuss below, this can be attributed to the fact that the correction term $\delta_2$ is partially canceled by
the binding energy of monopole polarons.

The degeneracy of the ice-I states is estimated following Pauling's method: $W_{z3} \sim 2^N \times (3/4)^{N_{z3}}$, where the factor $3/4$ is the fraction
of vertices satisfying the $Q=\pm 1$ ice rules, and $N_{z3} = 2N/5$ is the number of $z=3$ vertices. 
By imposing further constraints from the $z=4$ vertices, we estimate the degeneracy of the ice-II 
phase: $W_{z4} \sim (3/16)^{N_{z4}}\times W_{z3}$, where $N_{z4} = N/5$ is the number of even vertices.
Since these two crossovers are close to each other and to the low-$T$ charge-ordering transition, only slight washed-out shoulders can be seen
around the corresponding entropy densities in the entropy curve shown in  Fig.~\ref{fig:simulation}(a).

The residual charges $Q = \pm 1$ in the ice-II phase interact with each other through the magneitc Coulomb 
interaction $\mu_0 Q_\alpha Q_\beta/4\pi r_{\alpha\beta}$.  The Coulomb energy is minimized 
when adjacent $z=3$ vertices carry charges of opposite signs.
Consequently, the system undergoes a continuous transition at $T_c \approx 0.6 D$ into a phase with long-range ordering of magnetic charges. 
Numerically, the charge-ordering is signaled by a sharp peak in the $c(T)$ curve shown in Fig.~\ref{fig:simulation}(a). 
The resultant staggered charge order (an NaCl-type ionic crystal) shown in Fig.~\ref{fig:lattice}(b) 
resembles that observed in the kagome spin ice~\cite{moller09,chern11}. The broken sublattice $Z_2$
symmetry of the charge-ordered phase indicates that the corresponding transition belongs to the 2D Ising universality class.

Notably magnetic moments remain disordered in the charge-ordered ice-III phase, [green sector in Fig.~\ref{fig:simulation}(b)]
similar to the case of kagome spin ice~\cite{moller09,chern11,chern12}.
As shown in Fig.~\ref{fig:simulation}(a), the entropy density $s(T)$ exhibits a gentle step-like decrease to a nonzero value 
below the charge-ordering transition. We estimated a residual entropy $s_{\rm co} \approx 0.143$ per spin by numerically
simulating a toy model whose ground state is the degenerate charge-ordered ice phase. At very low temperatures,
corrections to the Coulomb's interactions induce a discontinuous transition below which the system develops a long-range
dipolar order. The perfectly ordered state shown in Fig.~\ref{fig:lattice}(c) has
a period-3 structure along the $x$ direction which shifts by one lattice constant when moving to the next row.
The corresponding order parameter $M$ exhibits a sudden increase as shown in  Fig.~\ref{fig:simulation}(b).

\begin{figure}
\includegraphics[width=0.9\columnwidth]{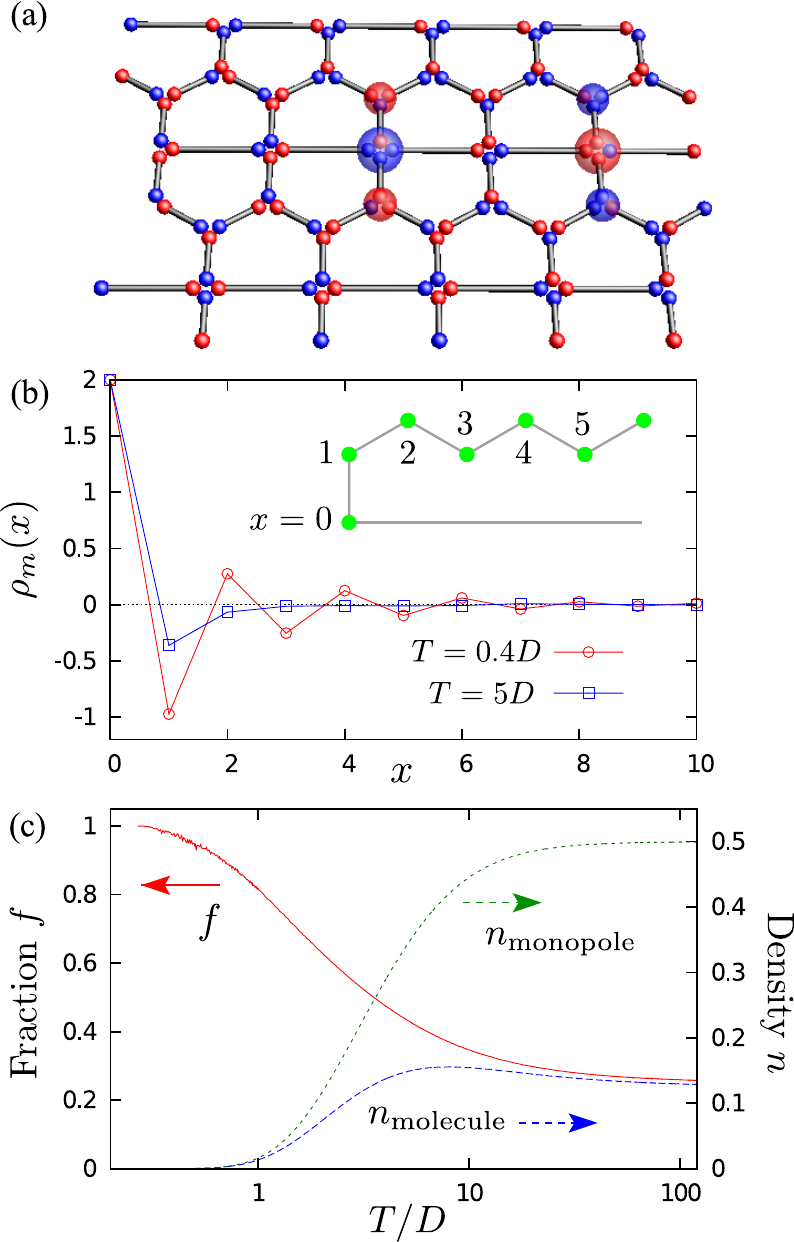}
\caption{\label{fig:monopole} (a) Schematic diagram showing  a pair of monopole molecules or polarons in the pentagonal lattice.
(b) Distribution of averaged magnetic charges $\rho_m(x) = \langle Q(x)\rangle$ around a $Q = 2$ monopole at two different temperatures.
Here $x = 0$ corresponds to the $Q=2$ site, and $x \ge 1$ denotes the position of $z=3$ vertices on the zigzag chain.
(c) Density $n$ (right axis) of magnetic monopoles,  monopole-molecules and their fraction $f = n_{\rm molecule}/n_{\rm monopole}$ (left axis) 
as a function of temperature.}
\end{figure}

The unique combination of even and odd vertices in the pentagonal lattice makes it possible
to study the intriguing interactions between the emergent $Q = \pm 2$ monopoles and the residual background charges at the $z=3$ sublattice.
In particular, the excitation of these emergent monopoles in the ice-I phase is stabilized as these topological defects are surrounded
by a cloud of net opposite charges through the magnetic Coulomb interaction, 
forming an entity similar to the electron polaron in crystalline lattice~\cite{landau33}. 
To demonstrate this unusual charge-charge correlation, Fig.~\ref{fig:monopole}(b)
shows the distribution of magnetic charges $\rho_m(x) = \langle Q(x) \rangle$ around a $Q = + 2$ monopole. 
The chrage density $\rho_m(x)$ is obtained numerically
by averaging over many independent monopole incidents from the Monte Carlo simulations of a large lattice ($L = 24$).
At higher temperatures, the monopole is indeed surrounded in a cloud of opposite charges on the zigzag chains 
with a correlation length of roughly two lattice constants. As $T$ is lowered, a NaCl-type staggered correlation starts to
develop along the zigzag chains, a precursor of the charge-ordering transition. More importantly, a strong correlation
can be seen between the $Q = 2$ monopole and its two nearest neighbors along the $y$ links.
This composite excitation resembles a linear triatomic molecule AB$_2$. A pair of such monopole molecule and the anti-molecule 
is shown in Fig.~\ref{fig:monopole}(a). It is worth noting that the formation of monopole molecules competes with
the staggered charge arrangement [Fig.~\ref{fig:lattice}(b)] favored by the Coulomb interactions between the residual charges at the $z=3$ sites.
As shown in Fig.~\ref{fig:simulation}(b), this competition also manifests itself in the slight suppression of the staggered charge order parameter $Q_s$ 
in the ice-I phase for a finite system .

The temperature dependence of the fraction $f$ of monopoles forming a AB$_2$-type molecule is shown 
in Fig.~\ref{fig:monopole}(c). In the high-$T$ paramagnetic phase with uncorrelated charges, roughly one quarter of the monopoles find themselves
surrounded by two opposite charges at the $z=3$ nearest neighbors. As temperature is lowered, this fraction increases significantly
and approaches 1 in the vicinity of the crossover temperature $T_{z4}$, indicating that almost all monopoles are in the molecule state.

The density of monopoles at $T \gtrsim T_{z4}$ is well approximated by $n_{Q=\pm 2} \sim 2 \exp(-\Delta_2/T)$~\cite{castelnovo11}.
Fitting our numerical result with this Arrhenius equation gives a monopole excitation energy $\Delta_2 \approx \varepsilon_2 - \varepsilon_0 \approx 4D$.
This is in stark contrast to the case of pyrochlore spin ice in which the monopoles acquire an additional excitation energy
due to the long-range part of the dipolar interactions~\cite{castelnovo08}.
The absence of a significant correction from the long-range dipolar terms can be attributed to the charge neutrality of
monopole molecules in the pentagonal ice. These composite quasiparticles are thus less sensitive to the long-range Coulomb corrections.
The result can also be understood as a cancellation due to the binding energy of monopole molecules.
The correction term coming from the long-range part can be estimated as $\delta_2 \sim 2\,[  2 v(r_{\rm nn}) - v(r_{\rm nnn}) ]$,
where $v(r) = (\mu_0/4\pi r)$, and $r_{\rm nn}$, $r_{\rm nnn}$ denote distances between a pair of nearest and next-nearest neighboring sites~\cite{castelnovo08}. 
On the other hand, by forming a AB$_2$ molecule, the monopole gains an energy $E_{\rm b} \sim (\mu_0/4\pi) (-2\times 2/r_{\rm nn} + 1/r_{\rm nnn})$,
which partially cancels the correction $\delta_2$.

\begin{figure}
\includegraphics[width=0.9\columnwidth]{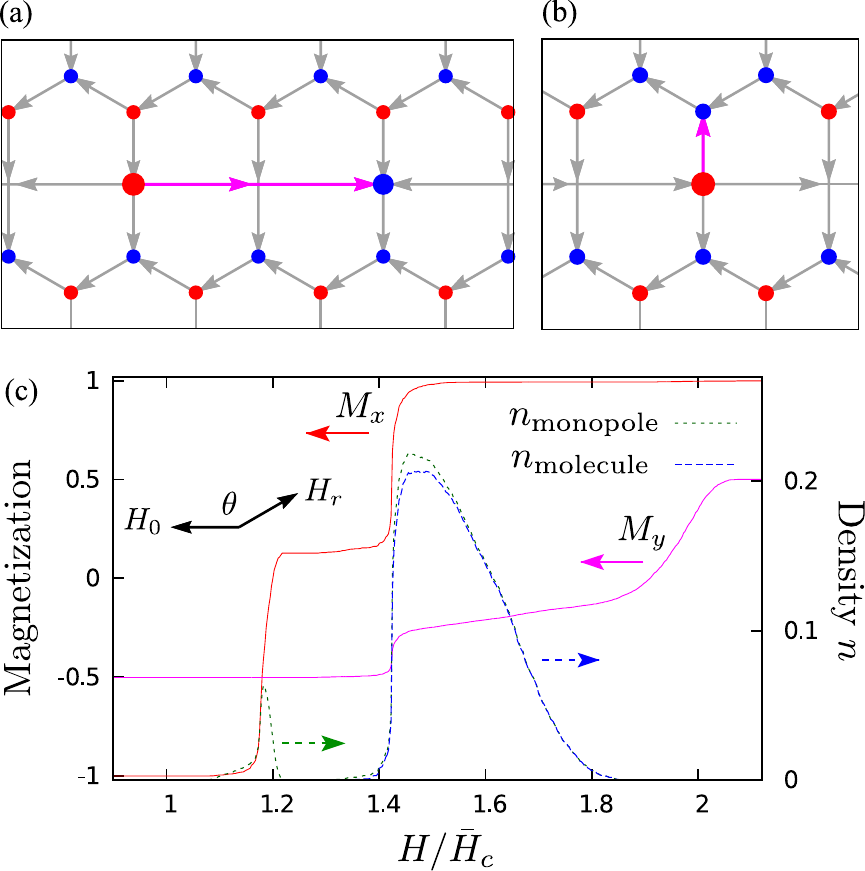}
\caption{\label{fig:m-reversal}(a) Creation of a pair of monopoles by flipping a $x$-link. (b) Inverting a $y$-spin
creates a monopole molecule in a charge-ordered background. (c) Magnetization (left) and density $n$ (right axis) of monopoles/monopole-molecules
as a function of the applied field. The reversal field is along a direction with an angle $\theta = 150^\circ$ relative to the polarization field direction.
The magnetizations are normalized with respect to the saturation $M_x$.}
\end{figure}

While the monopole polarons can likely be probed directly in a thermalized artificial pentagonal spin ice thanks to recent progress
in growth and thermal annealing technologies~\cite{morgan11a,cumings11,farhan13,gilbert13}, 
here we show that the monopole molecules also play an important role in the
magnetization reversal dynamics of the spin ice~\cite{ladak10,mengotti11,mellado10,morgan11}. 
We consider a pentagonal spin ice initially polarized by a strong magnetic field along the direction of $x$-links, say, $\mathbf H_0 \parallel -\hat{\mathbf x}$.
Subsequently, the field is switched off and a reversal magnetic field $\mathbf H_r$ is applied along the direction of $z$-links with an angle $\theta = 150^\circ$ 
from the initial field direction.
For simplicity, we have adopted the point-dipole approximation for the nanomagnetic islands~\cite{moller06,budrikis10,pollard12}.
A spin is inverted when the local field, which is a sum of the external field and the dipolar field coming from other spins, is greater than a critical $H_c$. 
This critical-field parameter varies slightly between links due to the unavoidable disorder present in the spin ice;
here we assume a uniform distribution of $H_c$ within the range ${\bar H}_c \pm \Delta H_c/2$ with $\Delta H_c = 0.1 {\bar H}_c$.

Fig.~\ref{fig:m-reversal} shows the simulated magnetization curves as a function of the reversal field.
The magnetization reversal occurs in two stages. At $H \approx {\bar H}_c/\cos(150^\circ)=1.15 {\bar H}_c$,
inversion of  a fraction of $x$-links creates monopole/anti-monopole pairs [illustrated in Fig.~\ref{fig:m-reversal}(a)],
giving rise to a step increase in $M_x$ and the first peak in the density of monopoles curve.  The pair creation events are largely independent
of each other. At $H \approx1.2 {\bar H}_c$ the remaining $x$ links flip to the right and therefore the density of monopoles decreases to zero. 
At a slightly larger field $H \approx 1.4 {\bar H}_c$, proliferation of monopoles mostly in the form of molecules results in a sudden
increase in both the $x$ and $y$ magnetizations. In a charge-ordered background due to the polarization field, creation of a single monopole
molecule can be achieved through the inversion of a single $y$-link as shown in Fig.~\ref{fig:m-reversal}(b). 
Above $H \approx 1.4 {\bar H}_c$ the inversion of the remaining $y$ links increases $M_y$ while reducing the density of molecules and monopoles coherently.
%As $M_y$ saturates, the density of monopole molecules saturates to zero as well. 

%In pentagonal spin ice, monopole polarons, composite quasiparticles made out of eight spins, are excitations to the ice-I phase.  The polaron transport in a highly ordered ionic crystals of magnetic charges is driven by an external magnetic field. Indeed, the proliferation of polarons, determine the dynamics of magnetization in pentagonal ice as shown by the results of magnetization reversal dynamics.  

The presence of composite quasiparticles in artificial spin ice is intriguing in itself and shows the rich physics of 
these unusual frustrated magnetic systems. Controlling and manipulating these magnetic composite quasiparticles
could open new avenues for the engineering of magnetronic circuitry, or magnetricity~\cite{bramwell09,giblin11}.
We hope our findings will encourage experiments aimed at realizing
artificial spin ices with mixed coordination numbers and the direct observation of the monopole polarons.

{\bf Acknowledgements}. We gratefully acknowledge insightful discussions with J. Cumings, %M. Gingras, P.~Holdsworth, R.~Moessner, 
C. Nisoli, P. Schiffer, and O.~Tchernyshyov.  G.W.C. thanks the support of the LANL Oppenheimer Fellowship. P.M. acknowledges the support from Fondecyt project No 11121397, and Conicyt, project No 79112004.

\end{document}